\documentclass[fleqn,12pt,twoside]{article}
\usepackage{espcrc1}

\usepackage{graphicx}
\newcommand{\be}{\begin{eqnarray}}
\newcommand{\ee}{\end{eqnarray}}
\def\lsim{\raise0.3ex\hbox{$<$\kern-0.75em\raise-1.1ex\hbox{$\sim$}}}
\def\gsim{\raise0.3ex\hbox{$>$\kern-0.75em\raise-1.1ex\hbox{$\sim$}}}

\def\b{{\bf b}}
\def\s{{\bf s}}

\def\Teff{{$T_{\rm eff}$}}
\newcommand{\AmS}{{\protect\the\textfont2
  A\kern-.1667em\lower.5ex\hbox{M}\kern-.125emS}}

\title{From quark-gluon plasma to hadron spectra}

\author{P.V. Ruuskanen\address{Department of Physics,
                 University of Jyv\"askyl\"a,\\
                 P.O. Box 35, FIN-40351 Jyv\"askyl\"a, Finland}}

\begin{document}

\maketitle

\begin{abstract}
Results on initial transverse energy production based on NLO perturbative QCD
calculation with final state saturation of produced minijets are used to
fix the initial energy density of produced matter. Assuming rapid
thermalization, this provides the initial conditions for a hydrodynamic
description of the expansion of final matter. Given a prescription of
the the decoupling of particles from the thermal system to free
particles, final transverse spectra of hadrons and integrated quantities
like multiplicity and transverse energy
can be calculated in the central rapidity region. Results are reported
                    and compared with measurements.
  \end{abstract}

%%%%%%%%%%%%%%%%%%%%5

\section{INTRODUCTION}

Conservative estimates from the measured transverse energy and particle
multiplicity at RHIC indicate that energy and particle densities well
above those of normal nuclear matter are formed in nuclear collisions.
Results from perturbative QCD (pQCD) calculation of minijet production
with final state saturation \cite{KJEskola} follow well the measured
multiplicities. However, the calculated transverse energy per unit
(pseudo)rapidity is larger by a factor $\sim 2.6$ than the measured
value. For achieving agreement between the calculation and the
experiment, the evolution of the produced matter must transfer a large
fraction of the energy from mid-rapidity towards the fragmentation
regions. In the hydrodynamic description the mechanism for the energy
transfer is the work done by the pressure in the expansion. Assuming
that initially the expansion is mainly in the longitudinal direction
leads to an asymmetry between the transverse and longitudinal directions
and to the desired energy transfer.

The work and the cooling of the matter during expansion will change the
momentum distributions both in transverse and the longitudinal
direction. In this work we assume that in the mid-rapidities the
longitudinal flow follows the scaling law, $v_z=z/t$ (collision takes
place at $t=0$ and $z=0$) and that we can ignore the small longitudinal
changes of densities. With these restricting assumptions we are not able
to address the problem of longitudinal momentum distributions but we can
calculate the final transverse momentum distributions when the
particles decouple at the end of the thermal stage. Calculation of
spectra, the folding of flow with thermal motion, is performed using the
Cooper and Fry \cite{CooperFry} prescription.

The justification for the use of hydrodynamics can be marginal for some
parts of the produced system but, first, it is clear that the bulk of
the matter is very dense leading to numerous secondary collisions and,
second, in describing the energy-momentum transfer due to these
collisions in terms of hydrodynamics the conservation laws are correctly
satisfied.

\section{INITIAL CONDITIONS FROM pQCD MINIJETS}

The perturbative QCD calculation of minijet production is a momentum
space calculation.  In order to define the initial densities a
connection between the momentum of the minijet and its space--time
formation point is needed.  At collider energies the hard partons of the
colliding nuclei are Lorentz contracted to a region of order
$2R_A/\gamma_{\rm cm}<< 1$ fm.  We consider the collision region as a
point in the longitudinal direction and assume that the rapidity of the
minijet coincides with the space--time rapidity of the formation point,
$y=\eta=(1/2)\ln[(t+z)/(t-z)]$.  The formation (proper) time we take to
be the inverse of the saturation scale, $\tau_0=1/p_{\rm sat}$.  Thus
the minijet matter forms along the hyperbola $t=\sqrt{z^2+\tau_0^2}$
with initial longitudinal flow velocity $v_z(\tau_0)=z/t$.

The basic quantities for the minijet production in a nucleon--nucleon
collision are
$
\sigma_{\rm jet}(p_{\rm sat},\sqrt s,\Delta y,A)\quad{\rm and}\quad
\sigma_{\rm jet}\langle E_T\rangle(p_{\rm sat},\sqrt s,\Delta y,A)\,,
$
the minijet cross section
and its first moment in transverse energy (momentum) for the
rapidity interval $\Delta y$, both integrated in
$p_T$ from  $p_T=p_{\rm sat}$ to infinity
\cite{KJEskola,Eskola:2000fc,Eskola:1988yh}. The number of minijets or
the transverse energy in $\Delta y$ in a nucleus--nucleus collision is
obtained by multiplying the corresponding cross section with the
nucleon--nucleon luminosity (with an extra factor 2 for the number of
minijets) which for a central collision is $T_{AB}(0)$, the
overlap function of transverse densities for the colliding nuclei:
$
T_{AB}(\b)=\int d^2\s T_A(|\b-\s|)T_B(s) =T_{AB}(b)\,,\
T_A(\s)=\int_{-\infty}^{+\infty} dz\rho_A(z,\s) = T_A(s)\,,
$
where $\b$ is the impact parameter and $\s$ the transverse coordinate.

Average densities are obtained by dividing with the volume $\Delta V=
\Delta zA_T=\tau_0\Delta y\,\pi R_A^2$. This procedure is easily
generalized to local density in the transverse plane of the collision.
The nucleon--nucleon luminosity for a transverse area element $d^2\s$ is
$T_A(|\b-\s|)T_{B}(s)$ and the volume element
$dV=dz\,d^2s=\tau \Delta y\,d^2s$ leading to \cite{ERRT}
$$
n_{\rm pQCD}(\tau_0,\s)=
{dN\over \tau_0dyd^2s}={1\over\tau_0\Delta y}
2T_A(|\b-\s|)T_{B}(s)\sigma_{\rm jet}\,,
%(p_0,\sqrt s,\Delta y,A)\,.
$$
and
$$
\epsilon_{\rm pQCD}(\tau_0,\s) =
{dE_T\over\tau_0dyd^2s} ={1\over\tau_0\Delta y} T_A(|\b-\s|)T_{B}(s)
\sigma_{\rm jet}\langle E_T\rangle\,.
%(p_{\rm sat},\sqrt s,\Delta y,A)\,.
$$

We next make a bold assumption of fast thermalization: we assume that
the thermalization time scale is the same as that of production and for
simplicity we take the system to be thermal right at the formation time
$\tau_0$. Two simple arguments can be given in favour of this
assumption. First, if we calculate the temperature either from
$\epsilon(\tau_0)$ or $n(\tau_0)$, the result is very closely the same.
This means that for thermalization only collisions causing energy
and momentum transfer are needed, number changing reactions are not
essential. Second, the time scale $\tau_0$ is associated with the
saturation scale and the formation time of more energetic (mini)jets
could be shorter so that some secondary interactions take place
already before the overall formation time.

%%%%%%%%%%%%%%%%%%%%% FIGURE %%%%%%%%%%%%%%%%%%%%%%%%%%%%%%%%
\begin{figure}[htb]
%\hspace{5mm}
\begin{minipage}[t]{70mm}
%\framebox[79mm]{\rule[-26mm]{0mm}{52mm}}
   \includegraphics[height=7.0cm]{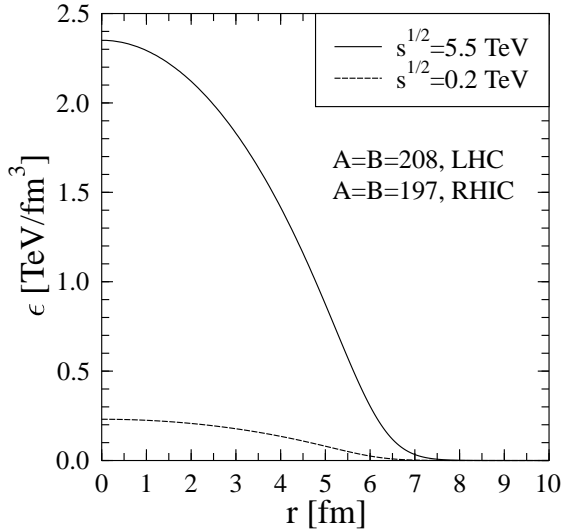}
%% \caption{Good sharp prints should be used and not (distorted)
%% photocopies.}
\end{minipage}
\hspace{\fill}
\begin{minipage}[t]{100mm}
 \vspace{-75mm}
 \begin{center}
  \begin{minipage}{65mm}
\caption{\protect\small
Transverse dependence of the initial energy
distribution for a gold-on-gold collision at RHIC (dashed line)
and lead-on-lead collision at LHC energy (solid line).  The average
value of the
saturation scale is $p_{\rm sat}=1.16$ GeV at RHIC and 2.03 GeV at LHC
with formation times 0.170 and 0.100 fm/c, respectively.}
  \end{minipage}
 \end{center}
\end{minipage}
\label{inicond}
\end{figure}
% \vspace{-10mm}
%%%%%%%%%%%%%%%%%%%%% FIGURE %%%%%%%%%%%%%%%%%%%%%%%%%%%%%%%%
%
In Figure 1 the transverse profile of the initial energy
distribution is shown for a gold-on-gold collision at RHIC (dashed line)
and lead-on-lead collision at LHC energy (solid line).  We comment on
the values of multiplicity and transverse energy of initial minijets
when discussing later the results on final hadron spectra.

\section{HYDRODYNAMICS OF EXPANSION}

We will assume isentropic expansion.  Studies of viscous effects
indicate that the entropy production during expansion does not produce
significant effects \cite{Chu:1986fu}.  The longitudinal flow is taken
to be boost invariant and to scale at central rapidity.  Then $v_z=z/t$
or equivalently $y=\eta$ and the energy density and transverse flow
velocity are of form $\epsilon=\epsilon(\tau,r)$ and $v_T=v_T(\tau,r)$
\cite{Bjorken:1983qr}. With cylindrical symmetry the equations reduce to
1+1--dimensional form and finding out the solutions numerically is
straightforward.

To close the set of hydrodynamical equations the equation of state (EoS)
is needed. We have assumed an ideal QGP at high temperatures and a
hadron gas including all hadrons and hadron resonances up to the mass
2~GeV. Repulsion of hadrons is described by mean field with $K=450$~MeV
and to induce a phase transition a bag constant is included in the EoS.
Calculations presented here are performed with the bag constant
$B^{1/4}=235$~MeV leading to value $T_c=165$~MeV for the phase
transition temperature \cite{Huovinen:1999tq,Sollfrank:1997hd}.

Finally, to turn the hydrodynamic quantities, the transverse velocity
$v_T(\tau,r)$ and the local temperature $T(\tau,r)$ to observable
quantities, the thermal motion, characterized by~$T$, must be folded
with the flow motion~$v_T$. This is done using the Cooper and Fry
prescription \cite{CooperFry}. Final particles are assumed to decouple
from the thermal
phase at a given density~$\epsilon_{\rm dec}$ or equivalently in the
zero baryon number case at temperature~$T_{\rm dec}$. The condition
$T(\tau,r)=T_{\rm dec}$ defines a hypersurface $\sigma^\mu$ and the
particle spectra are obtained as the net particle number flow through
this surface. For the boost-invariant cylindrically symmetric case the
integration can be reduced to one-dimensional integrals:
\be
 {dN\over dydp_T^2} &=& {g\over 2\pi}
  \sum_{n=1}^\infty (\pm 1)^{n+1} \int_\sigma  r\tau
  \big[-p_T I_1(n \gamma_rv_r{p_T\over T})
  K_0(n\gamma_r {m_T\over T})\,d\tau    \nonumber\\
  & & +m_T I_0(n \gamma_rv_r{p_T\over T})K_1(n\gamma_r{m_T\over T})\,dr
   \big]\,.
\label{eq:dec}
\ee
Without transverse flow the first term is missing and the second term
depends only on the transverse mass $M_T=\sqrt{p_T^2+M^2}$ and
temperature $T$.
Equation \ref{eq:dec} shows explicitly the breaking of $M_T$ scaling
through the $p_T$ dependent terms when $v_T\neq 0$. This is shown in
Figure 2 where pion, kaon and proton spectra are shown
for $v_T=0$ and $v_T=0.6$. The temperatures, $T=220$ MeV when $v_T=0$
and $T=120$ MeV when $v_T=0.6$, are so chosen that the slopes at large
$m_T$ are similar. For $v_T=0$ the mass shows up only in the starting
point of the spectrum. (There is a difference between bosons and
fermions from the alternating sign for the latter in the summation.)
For non-zero transverse flow the spectra deviate from the scaling
behaviour mainly in the region $p_T\lsim M$. When $p_T$ is clearly
bigger than $m$, the spectra become similar since $p_T\sim M_T$. In
nuclear collisions the spectra get contributions over a range of
transverse velocities, but the general features remain the same as in
Figure 2 for $v_T=0.6$.

%%%%%%%%%%%%%%%%%%%%% FIGURE %%%%%%%%%%%%%%%%%%%%%%%%%%%%%%%%
\begin{figure}[t!]
\hspace{5mm}
\begin{minipage}[t]{70mm}
%\framebox[79mm]{\rule[-26mm]{0mm}{52mm}}
   \includegraphics[height=7.5cm]{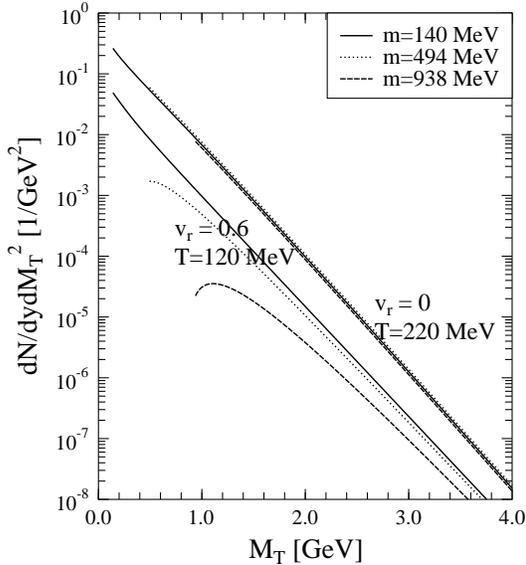}
%% \caption{Good sharp prints should be used and not (distorted)
%% photocopies.}
\label{mtscaling}
\end{minipage}
\hspace{\fill}
\begin{minipage}[t]{80mm}
\vspace{-70mm}
\begin{center}
\begin{minipage}{70mm}
\caption{\protect\small
Thermal spectra exhibit transverse mass scaling when $v_r=0$. For
non-zero
transverse velocity ($v_r=0.6$) transverse spectra of particles depend
not only on $M_T$ but also on $M$. Temperatures are arbitrary but so
chosen that all spectra have approximately the same slope at large
$M_T$.}
\end{minipage}
\end{center}
\end{minipage}
\end{figure}
%%%%%%%%%%%%%%%%%%%%% FIGURE %%%%%%%%%%%%%%%%%%%%%%%%%%%%%%%%

The buildup of transverse flow is one manifestation of the collective
behaviour of a dense system.  The estimates, using as input the Lorentz
contracted nuclear geometry and the observed multiplicities, give quite
high particle densities.  With such densities it would be difficult to
understand the dynamics of the produced matter if no collective effects
are observed since this would indicate that the final state particles
interact very weakly.

In any approach based on dominance of hard processes in the initial
particle production the predicted transverse energy will be consistent
with observations only if the system behaves collectively:  the
initially produced transverse energy is much higher than the observed
one as discussed in detail in Section 4. In the fast longitudinal
hydrodynamic expansion a large fraction of this energy is transferred
into the longitudinal motion which should show up in the broadening of
rapidity spectra from their initial shape.  In the present boost
invariant calculation only the energy loss in the central rapidity
region can calculated.

\section{RESULTS AND COMPARISON WITH EXPERIMENT}

The calculations are performed for central collisions when
the produced matter distributions are cylindrically symmetric.
Since the RHIC data is given for a 6~\%
centrality cut, we use in calculating the initial minijet production an
effective mass number for nucleons fixed to equal the average number of
participants for that centrality cut; for details
see~\cite{ERRT}.  We assume a similar centrality cut in
predictions for the LHC energies.  To show the magnitude of the change
from imposing the cut, we perform the calculations also with mass
numbers $A=197$ for gold at RHIC and $A=208$ for lead at LHC.

%%%%%%%%%%%%%%%%%%%%% FIGURE %%%%%%%%%%%%%%%%%%%%%%%%%%%%%%%%
\begin{figure}[htb]
%\hspace{5mm}
\begin{minipage}[t]{78mm}
   \includegraphics[height=7.5cm]{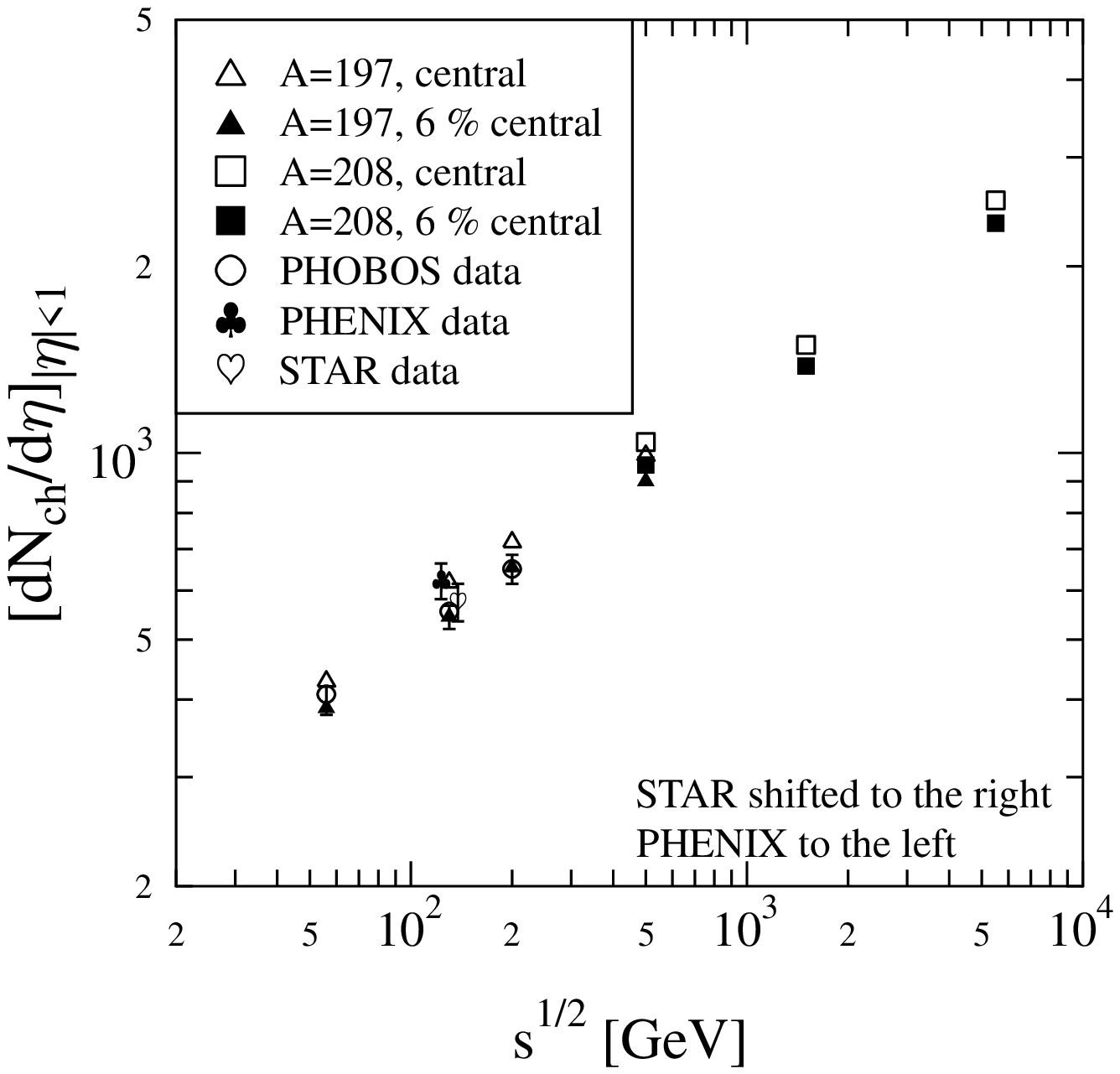}
\label{dnchdeta}
\vspace{-10mm}
\begin{center}
\begin{minipage}[h]{70mm}
\caption{\protect\small Charged particle multiplicity as function of
collision energy \cite{ERRT}. Calculations (triangles and squares) are
for the full
(open) and effective (filled) mass number determined from centrality.
}
\end{minipage} \end{center}
\end{minipage}
\hspace{\fill}
%
%\vspace{-75mm}
\begin{minipage}[t]{78mm}
   \includegraphics[height=7.5cm]{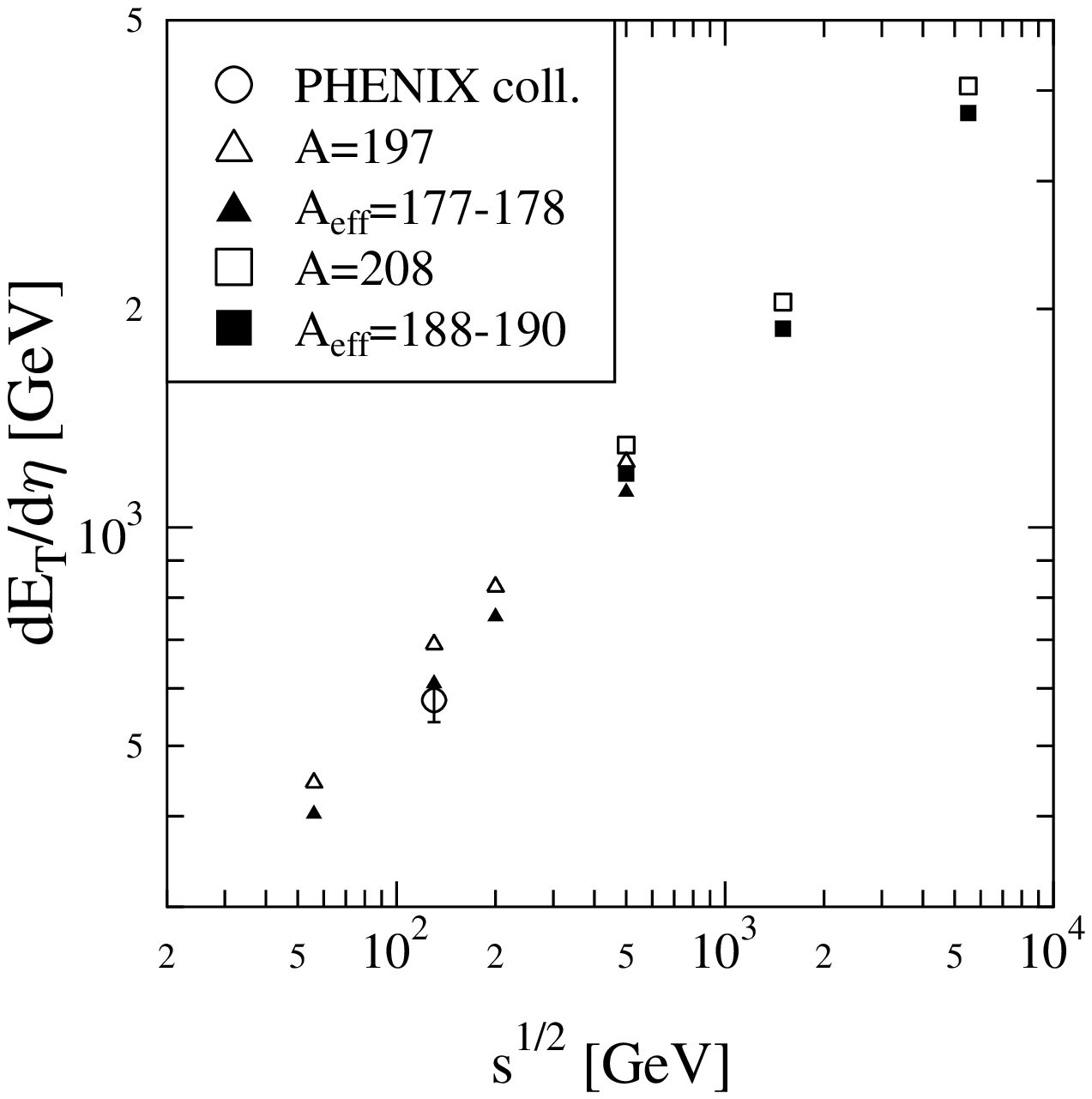}
\label{detdeta}
\vspace{-6.5mm}
\begin{center}
\begin{minipage}[h]{70mm}
\hspace{7mm}
\caption{\protect\small As in the left panel but for total
transverse energy \cite{ERRT}. The initial transverse energy at $\sqrt
s=130$ GeV is also displayed.}
\end{minipage} \end{center}
\end{minipage}
\end{figure}
%%%%%%%%%%%%%%%%%%%%% FIGURE %%%%%%%%%%%%%%%%%%%%%%%%%%%%%%%%

I will begin by presenting results on $p_T$--integrated quantities and
finish by discussing the transverse spectra. All results are for
mid-rapidity since the minijet calculation is performed at $y=0$
and boost invariance is assumed in the hydrodynamic calculation.

Figures 3 and 4 display the dependence
of the charged particle multiplicity and the total transverse energy on
the collision energy.  The charged particle multiplicity has been
measured by PHOBOS Collaboration \cite{PHOBOS} at three different
collision energies, $\sqrt{s_{NN}}=56,\ 130$ and 200 GeV.  There are
some uncertainties in the normalization of the minijet calculation.
E.g.,\ the saturation condition for the cut-off momentum in minijet
production, $p_{\rm sat}$, could contain extra factors of order 1 but
the simplest choice with no such factors is taken.  However, once the
normalization is fixed, the energy dependence is strongly constrained.
There are still uncertainties common to all hard scattering approaches
like those coming from higher order contributions.  For the minijet
calculation used here for the initial conditions, the contribution from
the NLO terms has been calculated for the transverse energy; for details
see \cite{KJEskola,EskolaT}.  The agreement of calculated
multiplicities
with the PHOBOS data \cite{PHOBOS} is very good both in normalization
and the energy
dependence. The values measured at $\sqrt s$ by PHENIX \cite{PHENIXDN}
and STAR \cite{STARDN} Collaborations are $5\ldots 10$ \% higher.

For transverse energy the value measured by PHENIX Collaboration at
$\sqrt s=130$ GeV \cite{PHENIXET}, 578 GeV, is $\sim 6$ \% smaller than
the calculated value 614 GeV.  This should be compared with the initial
value $\sim 1550$ GeV, also depicted in Figure 4 showing
an almost a factor of 3 reduction of $dE_T/d\eta$ due to the expansion.
On the other hand, as pointed out above, the multiplicity measured by
PHENIX \cite{PHENIXDN} is somewhat above our calculated value.  This
is an indication that, even though the energy in the transverse
degrees of freedom is reduced by a large factor during the expansion,
our calculated spectra, to be discussed below, are still slightly
shallower than the measured ones.

Figure 5 shows the results on effective temperatures
\Teff, the inverse slope of the spectrum, for pions, kaons and protons.
When the flow destroys the scaling in transverse mass, the slope of the
distributions becomes sensitive on the mass of the particle and changes
rapidly with $M_T$ (or $p_T$) especially in the region $M_T\lsim
(2\ldots 3)M$ as was seen in Figure 2.  For this
reason \Teff\ is displayed for two different transverse mass intervals,
$0.5<M_T\!-\!M<1.5$ and $2.5<M_T\!-\!M<3.5$, in the left and the right
panels of the Figure, respectively.  At the smaller transverse mass
values the mass dependence of slopes is very strong.

%%%%%%%%%%%%%%%%%%%%% FIGURE %%%%%%%%%%%%%%%%%%%%%%%%%%%%%%%%
\begin{figure}[t!]
%\hspace{5mm}
\begin{center}
   \includegraphics[height=8.5cm]{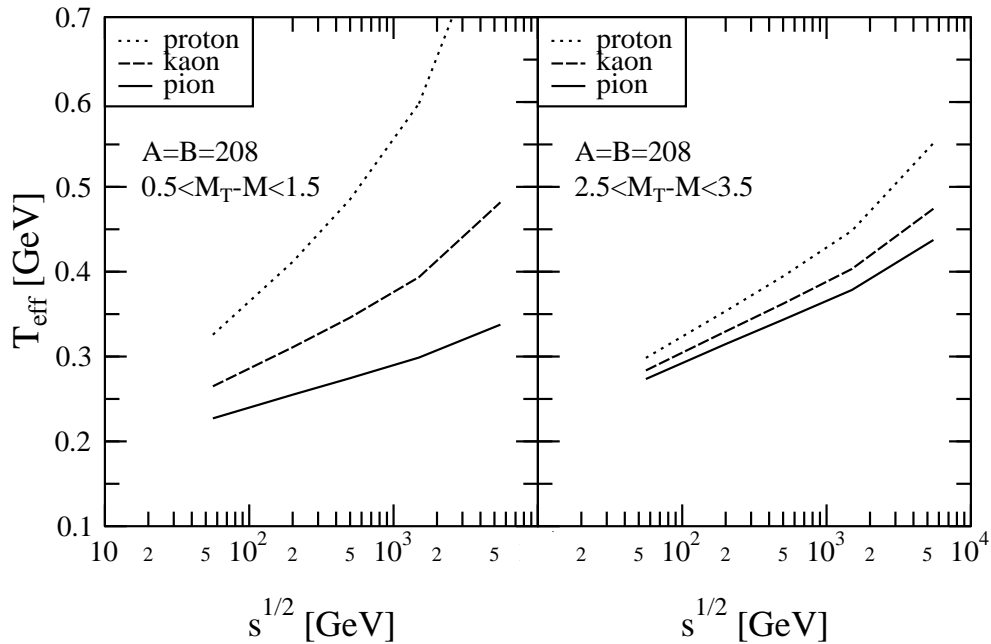}
\vspace{-4mm}
\begin{minipage}[h]{150mm}
\caption{\protect\small
Effective temperatures determined as inverse slopes of pion,
kaon and proton spectra at the LHC energy $\sqrt s=5500$ GeV for two
different transverse mass intervals. }
\end{minipage}
\end{center}
\label{Teffs}
\end{figure}
%%%%%%%%%%%%%%%%%%%%% FIGURE %%%%%%%%%%%%%%%%%%%%%%%%%%%%%%%%

At $\sqrt s= 130$ GeV, the STAR collaboration has data on inverse slopes
at small transverse mass, $0.05\,\,{\rm GeV}<M_T\!-\!M<0.45\,\,{\rm
GeV}$ \cite{JHarrisQM01}.  These are displayed in Table 1 (in
MeV units) with inverse slopes from the calculated spectra at $M_T\!-\!M
= 0.3$ GeV.  One observes that the calculated values are slightly larger
than the measured ones but the ratios of the measured or the
calculated slopes for different particles are quite similar.

%%%%%%%%%%%%%%%%%%%%% TABLE %%%%%%%%%%%%%%%%%%%%%%%%%%%%%%%%
\begin{table}[!htb]
\hspace{5mm}
\begin{minipage}[t]{70mm}
\begin{tabular}{|c|c|c|c|}
\hline
\Teff & Pion & Kaon & Proton \\
\hline
Measured  &190  &300  & 565 \\
Ratios  & 1  & 1.58 & 2.97 \\
Calculated & 238 & 387 & 652 \\
Ratios & 1  & 1.63  & 2.74 \\
\hline
\end{tabular}
\label{tableI}
\end{minipage}
\hspace{\fill}
\begin{minipage}[t]{80mm}
\vspace{-10mm}
\begin{center}
\begin{minipage}[t]{70mm}
\caption{\protect\small Inverse slopes, \Teff\ in MeV, of pion, kaon and
proton spectra measured at
RHIC at $\sqrt s=130$ GeV \cite{JHarrisQM01} and compared with
calculated values.}
\end{minipage}
\end{center}
\end{minipage}
\end{table}
%%%%%%%%%%%%%%%%%%%%% TABLE %%%%%%%%%%%%%%%%%%%%%%%%%%%%%%%%

Figure 6 shows the calculated average transverse momentum for
different particles.  In addition to the STAR measurement
\cite{JHarrisQM01} at RHIC, also the NA49~\cite{NA49} and the
UA1~\cite{UA1} data points are depicted.  Comparable to the STAR
measurement is $\langle p_T\rangle_{\rm all} = \sqrt{\langle
p_T^2\rangle}$ calculated for all particles and shown as the filled
circle.  The calculated value is somewhat too large as were also the
inverse slopes.  Comparison of the STAR measurement with the UA1 point
shows clearly a strong nuclear effect and the increase from the NA49
measurement indicates that the effect grows with energy.  In a thermal
model with expansion, this behaviour is understood in terms of
increasing collective flow as the initial densities grow with the
collision energy.

%%%%%%%%%%%%%%%%%%%%% FIGURE %%%%%%%%%%%%%%%%%%%%%%%%%%%%%%%%
\begin{figure}[h!]
\hspace{5mm}
\begin{minipage}[t]{60mm}
%\framebox[79mm]{\rule[-26mm]{0mm}{52mm}}
   \includegraphics[height=6.5cm]{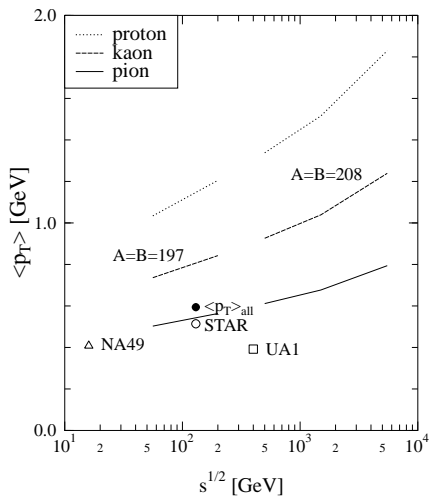}
\label{avpt}
\end{minipage}
\hspace{\fill}
\begin{minipage}[t]{90mm}
\vspace{-60mm}
\begin{center}
\begin{minipage}{80mm}
\caption{\protect\small The average $p_T$ of pions, kaons and nucleons
as a function of collision energy. In addition to the STAR
\cite{JHarrisQM01} data point, also NA49 \cite{NA49} and UA1
\cite{UA1} results are shown.}
\end{minipage}
\end{center}
\end{minipage}
\end{figure}
%%%%%%%%%%%%%%%%%%%%% FIGURE %%%%%%%%%%%%%%%%%%%%%%%%%%%%%%%%

Transverse spectra of pions, kaons and protons are shown in Figure~7
at $\sqrt s=200$~GeV and 5500 GeV.  The breaking of $M_T$
scaling is clearly seen.  It should be emphasized that the normalization
of kaons and protons relative to pions depends strongly on the
decoupling temperature which is here assumed to be the same for all
particles, $T_{\rm dec}\simeq 120$ MeV. It is more likely that the
number of kaons \cite{Kajantie:1986vc} and protons freezes out earlier
than at the kinetic freeze-out.  This would increase their normalization
relative to pions but the shape of the spectrum could change less since
the elastic collisions could still take place.  In calculating the
strange particle spectra, full chemical equilibrium is used at
freeze-out.  At lower energies strangeness is known to deviate from the
equilibrium values \cite{Sollfrank2}.  We have also ignored the non-zero
net baryon number which affects the normalization of (anti)proton
spectrum at RHIC but less at LHC.

%%%%%%%%%%%%%%%%%%%%% FIGURE %%%%%%%%%%%%%%%%%%%%%%%%%%%%%%%%
\begin{figure}[bht]
%\hspace{15mm}
\begin{center}
   \includegraphics[height=8.5cm]{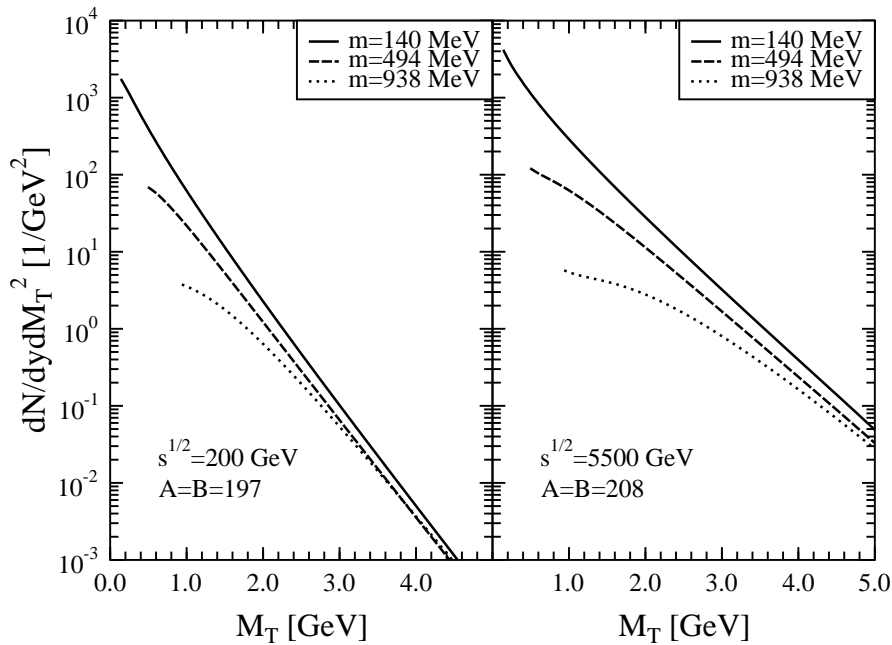}
\label{dndydmt}
\vspace{-4mm}
\begin{minipage}[h]{150mm}
\caption [1] {\protect\small Transverse mass distributions of pions,
kaons and protons.
In calculating the proton spectrum, zero baryon chemical potential is
used. The spectra contain also the particles from decays of resonances
up to $\Sigma(1385)$. }
\end{minipage}
\end{center}
\end{figure}
%%%%%%%%%%%%%%%%%%%%% FIGURE %%%%%%%%%%%%%%%%%%%%%%%%%%%%%%%%

%%%%%%%%%%%%%%%%%%%%% FIGURE %%%%%%%%%%%%%%%%%%%%%%%%%%%%%%%%
\begin{figure}[t!]
\hspace{5mm}
\begin{minipage}[t]{70mm}
%\framebox[79mm]{\rule[-26mm]{0mm}{52mm}}
   \includegraphics[height=7.5cm]{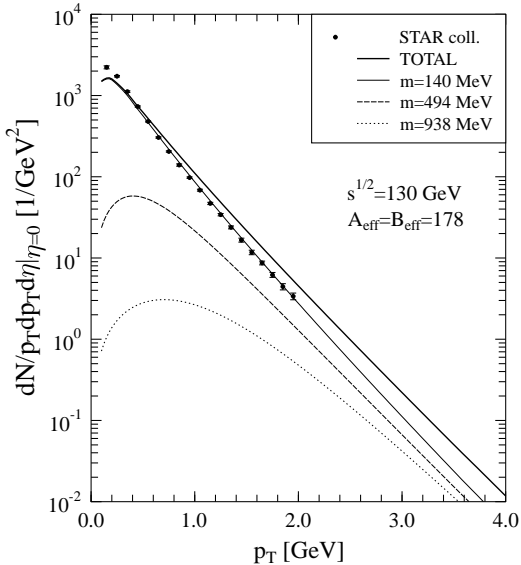}
%% \caption{Good sharp prints should be used and not (distorted)
%% photocopies.}
\label{pt130178}
\end{minipage}
\hspace{\fill}
\begin{minipage}[t]{80mm}
\vspace{-70mm}
\begin{center}
\begin{minipage}{70mm}
\caption{\protect\small
Transverse momentum spectrum of negative particles. The data points,
presented as filled circles are from STAR Collaboration
\cite{JHarrisQM01}. The calculated spectrum (thick solid line) contains
the contributions of $\pi^-,\ K^-$ and antiproton, shown also
separately.}
\end{minipage}
\end{center}
\end{minipage}
\end{figure}
%%%%%%%%%%%%%%%%%%%%% FIGURE %%%%%%%%%%%%%%%%%%%%%%%%%%%%%%%%

In Figure~8 the calculated transverse momentum spectrum of
negative particles at $\sqrt s=130$~GeV is compared with the spectrum
measured by the STAR Collaboration \cite{STARDN}.  There is an
indication of difference in shape at small transverse momenta.  In the
calculation the turning down of the spectrum comes from the Jacobian,
$p_T/M_T$.  At large values of transverse momentum the calculated
spectrum is somewhat above the measured one.  This was already seen when
comparing the values of inverse slopes and average transverse momenta.
In terms of the total reduction of energy in the transverse degrees of
freedom by almost a factor of three this is a small effect and we have
not tried to fine tune the calculation.

\section{CONCLUSIONS}

We have combined a minijet calculation of initial particle production at
collider energies with a hydrodynamic treatment for describing the
expansion of the produced matter. This allows us to predict particle
spectra, particle multiplicities and the transverse energy of final
particles as a function of the mass number of the colliding nuclei and
the collision energy.

The prediction from the minijet calculation of the initial particle
production which is most robust against details of the treatment of the
expansion is the multiplicity density, $dN/d\eta$ (or $dN/dy$), since
this assumes only an approximate conservation of entropy.  Since, as was
shown in Section 2, the energy and the number density of the initially
produced minijets correspond to the same thermal state, the number
changing reactions are not important for thermalization and the entropy
production after initial minijet production can be expected to be small.
The calculations agree very well with the PHOBOS data at each of the
three RHIC energies. At 130 GeV the PHENIX and STAR results are above
the calculated value but they are also higher than the PHOBOS result.

The initial energy per produced parton is quite large.  E.g.\ the
calculated average transverse momentum of initial minijets is $\langle
p_T\rangle_i\simeq 1.47$ GeV at the RHIC energy $\sqrt s=130$ GeV and
3.2 GeV at the full LHC energy $\sqrt s=5500$ GeV.  After the expansion
stage the calculated value at $\sqrt s=130$ GeV is $\langle p_T\rangle_f
= 0.594$ GeV to be compared with the measured value 0.514 GeV.  This
slight discrepancy in the predicted and measured transverse momentum
quantities shows up also in the inverse slopes or effective temperatures
determined from the spectra.  E.g., the calculated values of \Teff\ at
$M_T=0.3$ GeV are $20\ldots 25$ \% larger at $\sqrt s=130$ GeV than the
measured ones.  However, the predicted mass dependence comes out
qualitatively right.  The thermal model with collective transverse flow
predicts a strong dependence of the slope both on mass and transverse
mass in the transverse mass range $M_T-M\lsim 2M$.

The results from the minijet calculation with saturation assumption
combined with the hydrodynamic description of the expansion of final
matter look quite promising when compared with the first measurements at
RHIC.  This description of nuclear collisions provides a well
constrained framework for the study of other signals, like the photon
and lepton pair emission or the evolution of strangeness and the heavier
flavours during the final dense state.

\newpage
{\bf Acknowledgements:} I should like to thank my collaborators,
K.J.~Eskola, S.S.~R\"as\"anen and K.~Tuominen
for discussions and assistance and the Academy of Finland for financial
support.


\vspace{10mm}

\begin{thebibliography}{9}

\bibitem{KJEskola} K.J.~Eskola, this volume.

\bibitem{CooperFry} F.~Cooper and G.~Frye,
%``Comment On The Single Particle Distribution In The Hydrodynamic
%And Statistical Thermodynamic Models Of Multiparticle Production,''
Phys.\ Rev.\ D {\bf 10}, 186 (1974).
%%CITATION = PHRVA,D10,186;%%

\bibitem{Eskola:2000fc}  K.~J.~Eskola, K.~Kajantie, P.~V.~Ruuskanen and
K.~Tuominen,
%``Scaling of transverse energies and multiplicities with atomic number
% and energy in ultra-relativistic nuclear collisions,''
Nucl.\ Phys.\ B {\bf 570} (2000) 379, [arXiv:hep-ph/9909456].

\bibitem{Eskola:1988yh}
K.~J.~Eskola, K.~Kajantie and J.~Lindfors,
%``Quark And Gluon Production In High-Energy Nucleus-Nucleus Collisions,''
Nucl.\ Phys.\ B {\bf 323} (1989) 37.
%%CITATION = NUPHA,B323,37;%%

\bibitem{ERRT}
K.~J.~Eskola, P.~V.~Ruuskanen, S.~S.~R\"as\"anen and K.~Tuominen,
%   ``Multiplicities and transverse energies in central A A collisions
% at  RHIC and LHC from pQCD, saturation and hydrodynamics,''
Nucl.~Phys. A, in press, arXiv:hep-ph/0104010.
%%CITATION = HEP-PH 0104010;%%


\bibitem{Chu:1986fu} M.~C.~Chu,
%``One-Dimensional Hydrodynamics Of Ultra-relativistic Heavy Ion Collisions,''
Phys.\ Rev.\ D {\bf 34} (1986) 2764.

%\cite{Date:1989xi}
%\bibitem{Date:1989xi}
%S.~Date, M.~Mizutani, S.~Muroya and M.~Namiki,
%%``Hydrodynamical Expansion Of Viscous QGP Fluid With Phase
%%  Transition,''
%Prog.\ Theor.\ Phys.\  {\bf 82} (1989) 591.

%\cite{Bjorken:1983qr}
\bibitem{Bjorken:1983qr}  J.~D.~Bjorken,
%``Highly Relativistic Nucleus-Nucleus Collisions: The Central Rapidity Region,''
Phys.\ Rev.\ D {\bf 27}, 140 (1983).
%%CITATION = PHRVA,D27,140;%%


\bibitem{Huovinen:1999tq} P.~Huovinen, P.~V.~Ruuskanen and J.~Sollfrank,
%``Sensitivity of hadronic  and electromagnetic spectra to equation of
%state and initial energy density in the Pb + Pb collisions at SPS,''
Nucl.\ Phys.\ A {\bf 650} (1999) 227
[nucl-th/9807076].
%%CITATION = NUCL-TH 9807076;%%

\bibitem{Sollfrank:1997hd}
J.~Sollfrank, P.~Huovinen, M.~Kataja, P.~V.~Ruuskanen,  M.~Prakash and
R.~Venugopalan,
%``Hydrodynamical description of 200-A-GeV/c S + Au collisions:  Hadron
%and  electromagnetic spectra,''
Phys.\ Rev.\ C {\bf 55} (1997) 392
[nucl-th/9607029].

\bibitem{EskolaT}  % NLO laskut
K.~J.~Eskola and K.~Tuominen,
%``Production of transverse energy from minijets in
% next-to-leading order  perturbative QCD,''
Phys.\ Lett.\ B {\bf 489} (2000) 329
[hep-ph/0002008];
K.~J.~Eskola and K.~Tuominen,
% ``Transverse energy from minijets in ultrarelativistic nuclear
% collisions: A next-to-leading order analysis,''
Phys.\ Rev.\ D {\bf 63} (2001) 114006
[arXiv:hep-ph/0010319].

\bibitem{PHOBOS}
%\cite{Back:2000gw}
%\bibitem{Back:2000gw}
B.~B.~Back {\it et al.}  [PHOBOS Collaboration],
% ``Charged particle  multiplicity near mid-rapidity in central Au + Au
% collisions at s**(1/2) = 56-A/GeV and 130-A/GeV,''
Phys.\ Rev.\ Lett.\  {\bf 85} (2000) 3100
[arXiv:hep-ex/0007036];
%%CITATION = HEP-EX 0007036;%%
%\cite{Back:2001ae}
%\bibitem{Back:2001ae}
B.~B.~Back {\it et al.}  [PHOBOS Collaboration], arXiv:nucl-ex/0108009.
% ``Energy dependence of particle multiplicities in central Au + Au
%% collisions,''
%%CITATION = NUCL-EX 0108009;%%

\bibitem{PHENIXDN}  K.Adcox, {\it et al.} [PHENIX Collaboration],
 Phys.Rev.Lett. 86 (2001) 3500-3505 [nucl-ex/0012008].
% Centrality Dependence of Charged Particle Multiplicity in Au-Au
% Collisions at sqrt(s_NN)=130 GeV

\bibitem{STARDN}
C. Adler {\it et al.} [STAR Collaboration], Phys. Rev. Lett. 87 (2001)
112303  [nucl-ex/0106004].
  % tama on spektri + dN/deta_neg=280 paperi

\bibitem{PHENIXET}  K.Adcox, {\it et al.} [PHENIX Collaboration],
Phys.Rev.Lett. 87 (2001) 052301 [nucl-ex/0104015]

\bibitem{JHarrisQM01} J.~Harris for STAR Collaboration, Invited talk at
Quark Matter 2001.

\bibitem{NA49} H.~Appelsh\"auser {\it et al.} [NA49 Collaboration],
Phys.~Rev.~Letters, {\bf 82} (1999) 2471.

\bibitem{UA1}C.~Albajar {\it et al.}, Nucl.~Phys.~B~{\bf355} (1990) 261.

\bibitem{Kajantie:1986vc}
K.~Kajantie, M.~Kataja and P.~V.~Ruuskanen,
%``Strangeness Evolution In The Central Region Of A Heavy Ion
% Collision With Transverse Flow Effects,''
Phys.\ Lett.\ B {\bf 179} (1986) 153.

\bibitem{Sollfrank2}
%\cite{Sollfrank:1996hd}
J.~Sollfrank, P.~Huovinen, M.~Kataja, P.~V.~Ruuskanen, M.~Prakash and
R.~Venugopalan,
%``Hydrodynamical  description of 200-A-GeV/c S + Au collisions: Hadron
% and  electromagnetic spectra,''
Phys.\ Rev.\ C {\bf 55} (1997) 392
[arXiv:nucl-th/9607029].
%%CITATION = NUCL-TH 9607029;%%


\end{thebibliography}
\end{document}